# FeSe(*en*)$_{0.3}$ – Separated FeSe layers with stripe-type crystal structure by intercalation of neutral spacer molecules

Juliane Stahl and Dirk Johrendt[*]

Department Chemie, Ludwig-Maximilians-Universität München, Butenandtstr. 5-13 (D), 81377 München, Germany

**Solvothermal intercalation of ethylenediamine molecules into FeSe separates the layers by 1078 pm and creates a different stacking. FeSe(*en*)$_{0.3}$ is not superconducting although each layer exhibits the structure and Fermi surface of superconducting FeSe. FeSe(*en*)$_{0.3}$ requires electron-doping for high-$T_c$ like monolayers FeSe@SrTiO$_3$, whose much higher $T_c$ may arise from the oxide surface proximity.**

The most exciting discovery in the field of iron based superconductors during the last 5 years is probably the observation of superconductivity as high as 65 - 100 K in iron selenide (β-FeSe) monolayers grown on SrTiO$_3$ substrates with oxygen defects.[1-3] This has demonstrated the general potential of iron selenide layers to achieve superconductivity near or even above liquid nitrogen temperature, however, reasons for the giant increase of the transition temperature from 8 K in bulk FeSe are still under debate.[4] Calculations suggest an increased electron-phonon-coupling through the proximity of the substrate, which remains nevertheless too weak to explain a critical temperature of 65 K.[5] Interestingly, FeSe monolayers grown on defect-free SrTiO$_3$ or on graphene are not super-conducting,[6] while recent experiments with potassium-doped three-layer films suggest that high-$T_c$ superconductivity in FeSe requires electron doping of the layers.[7] This is in line with the fact that the $T_c$ of FeSe increases from 8 K to about 30 K through intercalation of alkaline metals.[8] Unfortunately, these materials are phase separated into a strongly magnetic non-superconducting phase and a superconducting phase of still unclear structure.[9] Relatively high transition temperatures up to 45 K occur in intercalation compounds of FeSe with organic molecules as spacers and alkaline metals as electron donors.[10-14] Consequently, neutrally intercalated FeSe with a large interlayer distance and weak interactions would serve as a bulk analogue of the undoped non-superconducting FeSe monolayers mentioned above without the proximity of the oxide surface. Given that the detailed structure of the monolayers are still lacking, the structure of such a "free monolayer" between weak interacting neutral molecules is a new piece in the unresolved puzzle of superconductivity in iron selenide.

In this communication we report the synthesis and crystal structure of iron selenide intercalated by ethylenediamine (*en*) molecules through a solvothermal route in an autoclave. Single crystal X-ray diffraction of a black plate-like crystal revealed a monoclinic *C*-centred lattice with $a \approx b \approx 387$ pm, $c = 2155$ pm and $\beta \approx 91°$. Relatively poor crystal quality and twinning impeded a satisfactory solution, therefore the structure was subsequently solved and refined from X-ray powder diffraction data. The atoms of the *en* molecules are not resolved and *en* was treated as a rigid body with geometry from literature.[15] The final Rietveld refinement and the resulting crystal structure are shown in Fig. 1, structure data are compiled in Tab. 1.

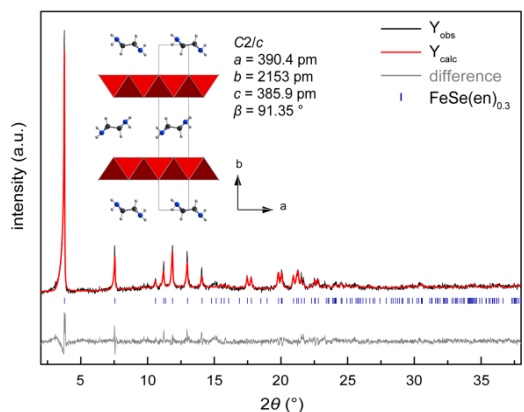

**Fig. 1.** Rietveld refinement of X-ray powder diffraction data (Mo K$_\alpha$) of FeSe(*en*)$_{0.3}$. The inset shows a projection of the structure with layers of edge-sharing FeSe$_{4/4}$ tetrahedra and intercalated *en* molecules..

*Corresponding author. Email: johrendt@lmu.de

**Tab. 1.** Structural parameters for FeSe(en)$_{0.3}$ from Rietveld refinement.

| atom | site | x | y | z | occ. | B$_{iso}$ |
|------|------|------|--------|--------|------|-----------|
| Fe1  | 4e   | 0.5    | 0.25   | 0.25   | 1    | 0.7       |
| Se1  | 4e   | 1      | 0.1822 | 0.25   | 1    | 0.7       |
| C1   | 8f   | 0.1928 | 0.9943 | 0.0325 | 0.3  | 0.7       |
| N1   | 8f   | 0.3902 | 0.0328 | 0.7938 | 0.3  | 0.7       |
| H1   | 8f   | 0.2617 | 0.0066 | 0.2991 | 0.3  | 0.7       |
| H2   | 8f   | 0.2507 | 0.9454 | 0.9887 | 0.3  | 0.7       |
| H3   | 8f   | 0.5543 | 0.0599 | 0.9468 | 0.3  | 0.7       |
| H4   | 8f   | 0.544  | 0.0026 | 0.6559 | 0.3  | 0.7       |

Space group: $C2/c$ (No. 15); $a$ = 390.37 pm, $b$ = 2152.7 pm, $c$ = 385.88 pm, $\beta$ = 91.35 °; $R_p$ = 3.31 %, $R_{wp}$ = 4.42 %; C, N and H occupancies are fixed to 0.3; B$_{iso}$ is fixed to 0.7 for all atoms;

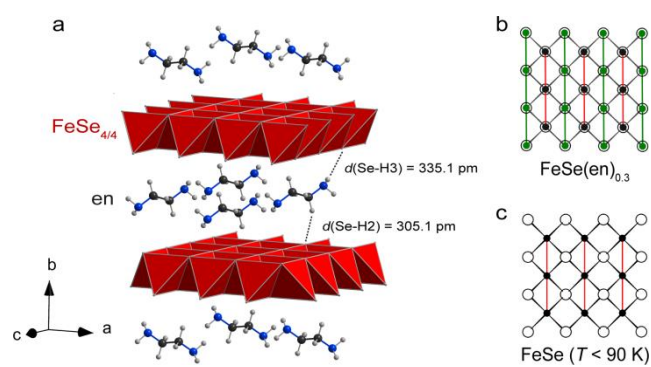

**Fig. 2. Fig. 2.** (a) Structure of FeSe(en)$_{0.3}$ (en partly omitted). (b,c) Projections of FeSe(en)$_{0.3}$ and FeSe layers with the shorter Fe-Fe bonds marked by red and green lines.

The amount of intercalated en was determined by chemical analysis. Deintercalation of en molecules starts near 200°C in argon atmosphere. Heating to 230°C completely removes en and FeSe is regained. The liquid section of the decomposed product was investigated by $^1$H and $^{13}$C NMR which showed pure en, thus no other molecular species had been intercalated. Thermogravimetric analysis confirmed the deintercalation and yielded a molar ratio of FeSe : en = 3 : 1. Energy-dispersive spectroscopy measurements (EDS) and inductively coupled plasma optical emission spectroscopy (ICP-OES) gave Fe : Se of 0.96 : 1 and 0.85 : 1, respectively. ICP-OES analysis yielded a ratio of FeSe : en = 3 : 1. CHN elemental analysis gave a C : H : N ratio of 2 : 9.21 : 1.70 which is not fully consistent with the formula of en $C_2H_8N_2$. The deviation is attributed to contamination with residues of the detergent.

The FeSe layers in monoclinic FeSe(en)$_{0.3}$ are separated by 1078 pm and stacked in a way different from β-FeSe (Fig. 2a). In the latter, iron and selenium atoms are stacked one above the other (Fig. 2c). Every second layer is shifted in the intercalated compound, where iron and selenium are now stacked alternatively (Fig. 2b). Similar stacking of layers is known from LaMnSi$_2$-type structures.[16] The FeSe$_{4/4}$ tetrahedra are weakly distorted with Fe-Se distances of 241.9 pm and 243.7 pm and angles ⊀Se-Fe-Se between 105.79° and 112.11° (FeSe: 238.2 pm, 104.3°-112.3°).[17] The Fe-Fe distances in the weakly distorted square Fe net are 271.2 and 277.7 pm and reveal the typical stripe-type motif of the shorter Fe-Fe bonds shown as red and green lines in Fig. 2. Thus the structures of the respective iron selenide layers are very similar in FeSe(en)$_{0.3}$ and pure orthorhombic FeSe (T < 90 K), except slightly longer Fe-Se bonds and some flattening of the tetrahedra in the intercalated compound.

The positions of the en molecules give shortest C-H···Se distances of 305.1 pm and N-H···Se distances of 335.1 pm (Fig. 2), comparable with the H-Se distances in ammonia intercalated FeSe.[10, 14] The distance of the FeSe layers in FeSe(en)$_{0.3}$ of 1078 pm agrees with those of superconducting $A_x$(en)$_y$Fe$_{2-z}$Se$_2$ (A = Li, Na) with 1037 pm to 1095 pm.[11-13] These compounds contain alkali ions in addition to en between the FeSe layers with a ThCr$_2$Si$_2$-type like stacking.

High-temperature PXRD data indicate an irreversible structural transition beginning at 180°C (Fig. 3 a) with a continuous decrease of the monoclinic angle till 200°C (inset in Fig. 3), where deintercalation of the en molecules starts. To further investigate the phase transition, samples of FeSe(en)$_{0.3}$ were prepared by intercalation of en into transport grown FeSe crystals (see notes).

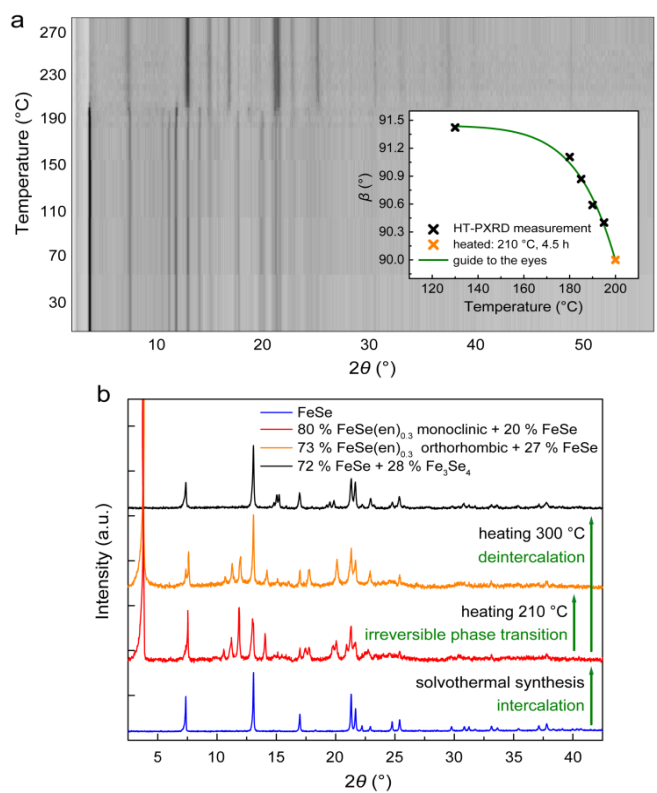

**Fig. 3.** a) Film plot of the in-situ high-temperature X-Ray powder diffraction data (Mo K$_\alpha$). Insert: Trend of the monoclinic angle β from Rietveld refinements. Additionally the angle β of the orthorhombic sample heated to 210 °C is inclosed (orange). b) X-Ray powder diffraction patterns of transport grown FeSe (blue), after intercalation of (en) via solvothermal method (red), product heated to 210 °C (orange) and 300 °C (black), respectively.

Intercalation of *en* into preformed FeSe in not complete under these conditions and the products contain some unreacted FeSe (red curve in Fig. 3 b). Heating of the monoclinic product to 210 °C for 4.5 h under argon atmosphere yields only orthorhombic FeSe(*en*)$_{0.3}$ ($\beta$ = 90°, space group *Cmcm*) with slightly increased amount of FeSe (orange curve in Fig. 3 b). This indicates the onset of the decomposition occurs simultaneously with the irreversible structural transition. We suggest that the transition is driven by the beginning deintercalation of *en*, which impedes further studies of the high temperature phase. Further heating to 300 °C leads to completely deintercalated FeSe and Fe$_3$Se$_4$ (back curve in Fig. 3 b) which is consistent with the *in-situ* high-temperature PXRD measurements.

The magnetic dc-susceptibility of FeSe(*en*)$_{0.3}$ at different fields shows continuously increasing paramagnetism over the whole temperature range (Fig. 4). Plots of the inverse susceptibilities were not linear. Isothermal magnetization curves and the field dependency of $\chi$ indicate a ferromagnetic component which may origin from traces of ferromagnetic impurities. In order to estimate the true paramagnetic susceptibility, the data were corrected by the Hondo-Owen procedure (green data points in Fig. 4), which is basically an extrapolation of the magnetization to infinite external field ($1/B \to 0$). This yields a significantly smaller susceptibility for FeSe(*en*)$_{0.3}$ of $2\times10^{-3}$ cm$^3$/mol at 295 K which nevertheless remains one order of magnitude larger than in FeSe which is $6.6\times10^{-4}$ cm$^3$/mol. The ac-susceptibility reveals no superconductivity in FeSe(*en*)$_{0.3}$ above 3 K. The left inset in Fig. 4 shows the data in comparison with transport grown superconducting FeSe. The absence of superconductivity in the undoped layer is expected and in line with the observations about FeSe monolayers on defect-free SrTiO$_3$ or graphene.[6]

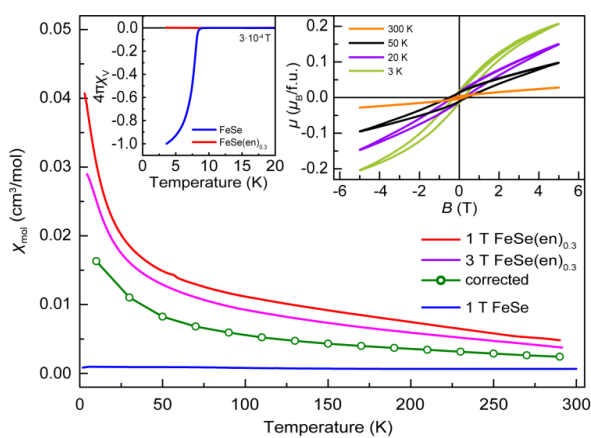

**Fig.4.** Magnetic dc-susceptibility, isothermal magnetisation and low temperature ac-susceptibility of FeSe(*en*)$_{0.3}$ and transport grown FeSe for comparison.

One may argue that the different stacking of the layers is responsible for the absence of superconductivity and not the lack of electron doping. We have calculated the Fermi surfaces of orthorhombic $\beta$-FeSe, FeSe(*en*)$_{0.3}$ (*en* molecules were omitted in the calculation) and hypothetically electron-doped FeSe(*en*)$_{0.3}$ shown in Fig 5. FeSe(*en*)$_{0.3}$ largely retains the typical Fermi surface topology of $\beta$-FeSe in spite of the different layer stacking, whereby the 2-dimensional character gets more pronounced due to the much bigger layer separation. Adding about 0.2 electrons per FeSe(*en*)$_{0.3}$ increases the Fermi energy and the hole-like parts of the surface around the $\Gamma$-point vanish. This is exactly what has been observed in three-layer FeSe which becomes superconducting only by doping with potassium.[4,7]

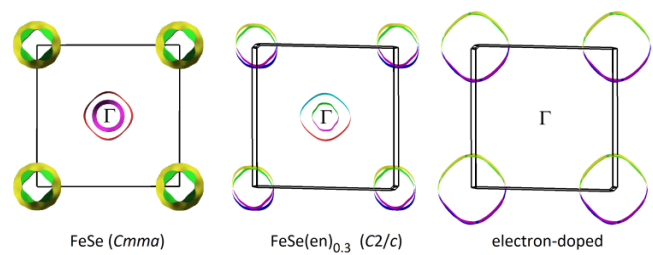

**Fig.5.** Fermi surfaces of $\beta$-FeSe (*Cmma*), FeSe(*en*)$_{0.3}$ (*C2/c*, *en*-molecules omitted) and hypothetically electron-doped FeSe(*en*)$_{0.3}$

In conclusion we have shown that the intercalation of a remarkably small amount of ethylenediamine molecules between FeSe layers increases the layer spacing to 1087 pm in FeSe(*en*)$_{0.3}$. Thus we have realized very weakly interacting and charge neutral FeSe layers with a structure almost identical to those of superconducting FeSe. We consider FeSe(*en*)$_{0.3}$ as a bulk analogue to the monolayer materials grown on SrTiO$_3$ albeit without the proximity of the rigid oxide surface. Our results support recent findings that monolayers require electron doping to become superconducting at high $T_c$ and we show evidence that this is also the case for *en*-intercalated bulk materials. The latter have so far reached critical temperatures of 45 K which is well below 65-100 K of the monolayers. One possible reason may be the additional increase of the electron-phonon coupling in the monolayers[5] on the rigid oxide substrate in contrast to the rather soft bearing of the FeSe layers between *en*-molecules.

## Notes and references

X-ray diffraction patterns were collected using a Stoe Stadi P diffractometer (Mo-$K_{\alpha1}$ radiation (70.93 pm); Ge-111 monochromator) with capillary sample holder. TOPAS[19] was used for Rietveld refinements. Single crystal analysis was performed on a Bruker D8-Quest diffractometer (Mo-Kα1 radiation (70.93 pm); graphite monochromator). Compositions of the samples regarding the Fe : Se ratio were investigated by energy-dispersive spectroscopy measurements (EDS) on a Zeiss Evo-Ma10 microscope with Bruker X-Flash 410-M detector and by inductively coupled plasma optical emission spectroscopy (ICP-OES). For analysis of the C : N : H ratio CHNS elemental analysis was used. Thermogravimetric analysis was used to track and quantify the deintercalation of *en*. The deintercalated section was investigated by $^1$H and $^{13}$C NMR. Magnetic measurements were carried out using a custom-made ac-susceptometer and a commercial MPMS XL SQUID dc-magnetometer (Quantum Design, San Diego, USA).